\documentclass[twocolumn,superscriptaddress,secnumarabic,amssymb, nobibnotes, aps, bm,pre,floatfix,longbibliography]{revtex4-2} 
\usepackage{amsmath,amssymb}
\usepackage{bm}
\usepackage{tipa}
\usepackage{upgreek}
\usepackage{comment}
\usepackage{mathrsfs}
\usepackage{graphicx}
\usepackage{braket}
\usepackage{enumitem}
\usepackage{natbib}
\usepackage{mathbbol}
\usepackage{booktabs}
\usepackage{gensymb}
\usepackage[normalem]{ulem}
\usepackage{color}
\usepackage[colorlinks,bookmarks=true,citecolor=blue,linkcolor=red,urlcolor=blue]{hyperref}

\usepackage{pifont}

\begin{document}

	\title{Successive electron-vortex binding in quantum Hall bilayers at $\nu=\frac{1}{4}+\frac{3}{4}$}

	\author{Glenn Wagner}
	\affiliation{Department of Physics, University of Zurich, Winterthurerstrasse 190, 8057 Zurich, Switzerland}

    \author{Dung X. Nguyen}
\affiliation{Center for Theoretical Physics of Complex Systems, Institute for Basic Science (IBS), 34126 Daejeon, Korea}
\affiliation{Basic Science Program, Korea University of Science and Technology (UST), 34113 Daejeon, Korea}

	\begin{abstract}
    Electrons in a quantum Hall fluid can bind with an integer number of vortices to form composite fermions and composite bosons. We show that the quantum Hall bilayer at filling $\nu=\frac{1}{4}+\frac{3}{4}$ with interlayer separation $d$ can be well-described in terms of these composite particles. At small $d$ the system can be understood as interlayer paired electrons and holes, whereas at large $d$ the system is best understood in terms of composite fermions with four vortices attached to each electron. By computing the overlaps of trial wavefunctions with the ground state from exact diagonalization, we find that as $d$ increases, the number of vortices that attach to each electron increases. We also construct trial states for two types of excitation, the Goldstone mode and a meron excitation. These two trial states have good overlaps with the lowest excited states in the exact diagonalization spectrum for small and intermediate $d$ respectively. 
	\end{abstract}
	
 	\maketitle

\section{Introduction}

Quantum Hall bilayers, i.e.~two layers of quantum Hall fluid separated by a distance $d$, offer a versatile platform where electron-electron interactions can be tuned \emph{in-situ} by changing the effective interlayer separation. Both experimental and theoretical effort spanning several decades has been devoted to this problem \cite{Eisenstein2004}. There are two length scales in the problem, the separation $d$ between the two layers of quantum Hall fluid and the magnetic length $\ell_B=1/\sqrt{eB}$ which is the typical extent of each electron's Landau orbital in a magnetic field $B$. By tuning the ratio of the interlayer separation $d$ and the magnetic length $\ell_B$ one can tune the system between the limit of two decoupled layers ($d/\ell_B\to\infty$) and the $SU(2)$ symmetric limit ($d/\ell_B=0$), where layer plays the role of pseudospin. In practice in experiments one tunes the magnetic length $\ell_B$ while keeping $d$ and the layer's filling factors $\nu_s=2\pi n_s\ell_B^2$ fixed, where $s=\uparrow, \downarrow$ is the layer pseudopsin and $n_s$ is the electron density in layer $s$. 

One of the most successful descriptions of the single layer quantum Hall effect involves composite particles formed by attaching an integer number $p$ of flux quanta to the electrons. For $p$ even the resulting particle is a composite fermion ($^p$CF) whereas for $p$ odd due to the statistical transmutation from the Aharonov-Bohm phase the resulting particle is a composite boson ($^p$CB) \cite{WilczekCF,Halperin111,CF,Girvin1987,Ye2008}. At filling $\nu=1/p$, the composite particles formed by attaching $p$ flux quanta experience no net magnetic field. States at filling factor $\nu=1/p$ can thus be thought of as a Fermi liquid of CFs for even $p$ \cite{FradkinLopez,HLR} and as a Bose condensate of CBs for odd $p$ \cite{Zhang1989,Read1989}. 

In the quantum Hall bilayer with the filling fraction fixed to $(\nu_\uparrow,\nu_\downarrow)=(\frac{1}{4},\frac{3}{4})$, the composite particle description can help understand the limit of decoupled layers. In the top layer we can attach four flux quanta to each electron to form $^4$CFs that experience no net magnetic field. In the bottom layer we can attach four flux quanta to the holes to form anti-$^4$CFs. In the large $d$ limit, the layers form two independent composite Fermi liquids: A Fermi liquid of $^4$CFs in the top layer and a Fermi liquid of anti-$^4$CFs in the bottom layer. The state in the opposite limit of small interlayer separation is also well-understood. In the $SU(2)$ symmetric limit, the system forms a quantum Hall ferromagnet \cite{Moon_Review,Ezawa_2009}, or equivalently an exciton condensate of electron-hole pairs, the so-called Halperin (111) state \cite{Halperin111}. The question is then how to describe the system at the intermediate distances between the two limits.

Exact diagonalization studies provide the ground state of the quantum Hall bilayer at different $d$ \cite{ED1,ED3,ED0,Park1,Park2,papicThesis,Wagner2021,Simon1,Simon2,Simon3,Milovanovic} and trial wavefunctions may be used to interpret this ground state. In fact, trial wavefunctions describing pairs of $^2$CFs and anti-$^2$CFs have proved successful for describing quantum Hall bilayers at $(\nu_\uparrow,\nu_\downarrow)=(\frac{1}{2},\frac{1}{2})$ for any interlayer separation \cite{Wagner2021,Hu2024,Simon2,Simon3}. The bilayer can be described as undergoing a BEC-BCS crossover between decoupled composite Fermi liquids at large $d/\ell_B$ and the exciton condensate at $d/\ell_B=0$ \cite{Hu2024,Wagner2021,Sodemann,Halperin2020,ruegg2024dualities}. At large $d$, the $^2$CFs do not experience a net magnetic field and form two independent composite Fermi liquids. However, there is an instability of the Fermi surface to interlayer BCS pairing \cite{Bonesteel,Pwave2,Cipri_thesis,CipriBonesteel,p_wave,Ruegg2023,lotric2024chernsimons} and as $d$ decreases, the pairing becomes stronger until the system enters the BEC regime with tightly bound $^2$CF/anti-$^2$CF pairs. In the tightly bound limit, the phases associated with the vortices from the $^2$CF and anti-$^2$CF cancel, leading to the electron-hole exciton condensate, or 111 state \cite{Halperin2020}. Trial wavefunctions on the sphere describing paired composite fermions indeed have large overlaps with the exact diagonalization ground state for the entire range of interlayer separations $d$ \cite{Hu2024,Wagner2021}. On the other hand, a trial wavefunction based on $^1$CBs has been proposed for the $(\nu_\uparrow,\nu_\downarrow)=(\frac{1}{2},\frac{1}{2})$ bilayer at intermediate distances $d\sim\ell_B$ \cite{ShouCheng}. Small-scale exact diagonalization in the planar geometry shows large overlaps with this trial state at intermediate distances. 

While the composite particle description has thus been successfully extended from single layer to bilayer quantum Hall systems at $(\nu_\uparrow,\nu_\downarrow)=(\frac{1}{2},\frac{1}{2})$, it remained unclear how well this description works for imbalanced layers, which have also been realized experimentally. Based on the results at $(\nu_\uparrow,\nu_\downarrow)=(\frac{1}{2},\frac{1}{2})$, we apply similar trial wavefunctions based on paired composite particles to the less well-studied imbalanced quantum Hall bilayer at filling $(\nu_\uparrow,\nu_\downarrow)=(\frac{1}{4},\frac{3}{4})$. In section \ref{sec:trial_wavefunction} we introduce a trial wavefunction for the spherical geometry that describes interlayer pairing of composite particles with $p$ fluxes attached. We compute the overlap of these trial wavefunctions with the exact diagonalization ground state in section \ref{sec:results}, which shows that the ideal number of fluxes attached $p$ increases from zero to four as $d$ is increased. In section \ref{sec:excitations} we also construct trial states for two branches of the excited states. We close with a discussion of the results and outlook towards future directions in section \ref{sec:discussion}.

\section{Ground state trial wavefunction}
\label{sec:trial_wavefunction}

We can write down a trial wavefunction describing $(\nu_\uparrow,\nu_\downarrow)=(\frac{1}{4},\frac{3}{4})$, motivated by similar wavefunctions at $(\nu_\uparrow,\nu_\downarrow)=(\frac{1}{2},\frac{1}{2})$ in \cite{Hu2024,Wagner2021}. Such a wavefunction for the disk geometry was written down in Eq.~(8) of Ref.~\cite{ShouCheng}. To construct the trial state, we first particle-hole transform the bottom layer, such that the system is described as electrons at $\nu=\frac{1}{4}$ filling in the top layer and holes at $\nu=\frac{1}{4}$ filling in the bottom layer. We can then attach $p$ fluxes to each electron in the top layer to form a $^p$CF/$^p$CB and attach $p$ fluxes to each hole in the bottom layer to form an anti-$^p$CF/anti-$^p$CB. We work on the sphere with $N_{\uparrow}$ ($N_{\downarrow}$) electrons in the top (bottom) layer and consider the sector where the number of flux quanta is $N_\phi=4(N_\uparrow -1)$. We work in the sector containing the 111 state which has $N_{\uparrow}+N_{\downarrow}=N_\phi+1$. The $^p$CFs/$^p$CBs feel a net flux $2q=N_\phi-p(N_{\uparrow}-1)=(4-p)(N_\uparrow -1)$.

The trial wavefunction describing pairing between $^p$CFs/$^p$CBs in the top layer and anti-$^p$CFs/anti-$^p$CBs in the bottom layer can be written in the spherical geometry as
\begin{eqnarray}
    \psi_{p}(\alpha) &=& \mathcal{P}_\mathrm{LLL} \prod_{i<j}(\Omega_i-\Omega_j)^p(\varpi_i-\varpi_j)^{*p}f(G)  \nonumber \\
    G(\Omega_i, \varpi_j)  &=& \sum_{n=0}^{N_\textrm{max}}\sum_{m=-q-n}^{q+n}  g_n Y_{qnm}(\Omega_i)  Y^*_{qnm}(\varpi_j),  \label{eq:trialwfmaintext}
\end{eqnarray}
where $f(G)=\mathrm{det}(G)$ for even $p$ and $f(G)=\mathrm{perm}(G)$ for odd $p$. $G$ stands for the matrix $G_{ij}=G(\Omega_i, \varpi_j)$. perm($G$) is the permanent of the matrix $G$, which is the symmetric version of the determinant as appropriate for bosons. 
$\Omega_i=(\theta_i,\varphi_i)$ is the spinor coordinate of electron $i$ in the top layer ($i=1,\dots,N_\downarrow$) and $\varpi_i$ is the spinor coordinate of hole $i$ in the bottom layer ($i=1,\dots,N_\downarrow$). The factor $(\Omega_i-\Omega_j)^p$ stands for the Jastrow factor which attaches $p$ flux quanta to each electron in the top layer, turning it into a composite particle. The factor $(\varpi_i-\varpi_j)^{*p}$ stands for the Jastrow factor which attaches $p$ flux quanta to each hole in the bottom layer turning it into an anti composite particle. $Y_{qnm}$ are the monopole harmonics on the sphere, which are the single-particle orbitals for a (composite) particle on a sphere in the presence of  $2q$ flux quanta. The orbitals are enumerated by Landau level index $n$ and the $L_z$ angular momentum quantum number $m$. $\mathcal{P}_\mathrm{LLL} $ describes the projection to the lowest Landau level, see Appendix for details of the projection procedure. 

The variational parameters $g_n$ describe the pairing between the composite particles. Their number is truncated at $N_\textrm{max}$ and in the current work we set $N_\textrm{max}=N_\uparrow$ Refs.~\cite{Wagner2021,Hu2024} have shown that high overlaps can be obtained with this number of variational parameters in the case $p=2$. We checked in the current work that including more variational parameters does not improve the overlaps significantly. Ref.~\cite{Hu2024} further proposed the ansatz $g_n=e^{\alpha n}$ such that $\alpha$ is the only variational parameter. This is a useful parametrization that allows us to recover the two extreme limits: $\alpha\gg1$ corresponds to occupying orbitals with large $n$, which leads to strong interlayer pairing, since correlations with large momenta correspond to tight pairing in real space. The opposite limit $\alpha\ll-1$ corresponds to filling only the lowest orbitals. This is the Fermi liquid (Bose condensate) limit with weak interlayer correlations for $p$ even (odd). 

\section{Overlaps with ground state trial wavefunction}
\label{sec:results}

\begin{figure*}
    \centering
    \includegraphics[width=\textwidth]{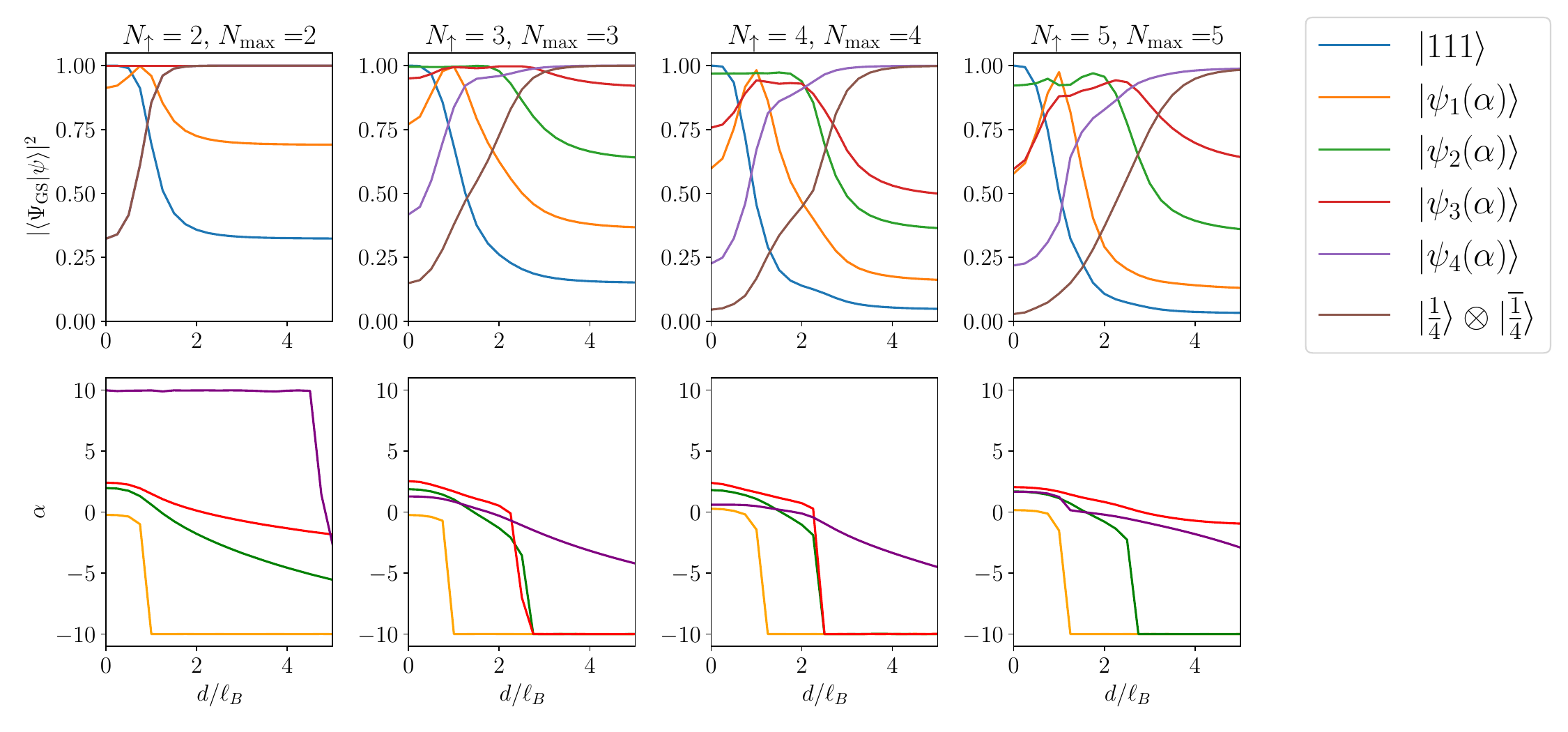}
    \caption{\textbf{Overlaps with ground state trial wavefunctions.} Top row: We show the overlap of the exact diagonalization ground state with the different trial wavefunctions $|\psi_p(\alpha)\rangle$. The trial wavefunctions have different numbers of vortices $p$ attached to each electron and hence describe paired composite bosons (odd $p$) or composite fermions (even $p$). As $d$ increases, the wavefunctions with more vortices attached perform better. For comparison, we also show the overlap with the Halperin (111) state, as well as the state describing decoupled composite Fermi liquids of $^4$CFs. Bottom row: The variational parameter $\alpha$ for the overlap-optimized trial wavefunction is shown, where we let $\alpha$ vary in the range $-10<\alpha<10$ for the optimization. $\alpha\gg1$ corresponds to strong interlayer pairing and occupation of higher angular momentum orbitals. $\alpha\ll-1$ corresponds to the Fermi liquid/Bose condensate limit. As $d$ increases, $\alpha$ tends to monotonically decrease, corresponding to weaker pairing between the layers. }
    \label{fig:overlaps}
\end{figure*}

In Fig.~\ref{fig:overlaps} we compare the trial wavefunction $|\psi_p(\alpha)\rangle$ with the exact diagonalization ground state $|\psi_\textrm{GS}\rangle$. We compute the overlaps of the wavefunctions for each interlayer separation $d$ in real space using Monte-Carlo integration and use a ``dual annealing" global optimization algorithm to optimize the overlaps over the variational parameter $\alpha$. Such an algorithm performs better than a gradient descent algorithm for finding the global optimum. We show the results of the optimized overlaps (top row) as well as the variational parameter $\alpha$ (bottom row) for four different system sizes. The Hilbert space dimension in the four cases is dim($L_z=0$)=16, 500, 21.773, 1.119.032 for $N_\uparrow=2,3,4,5$ respectively. For the case $p=0$, our trial wavefunction reduces to the Halperin 111 state with no free variational parameters since all terms with $n>0$ vanish upon lowest Landau level projection. This is the exact ground state at $d=0$. For the case $p=4$ and $\alpha\to-\infty$, our trial state reduces to the decoupled composite Fermi liquids of $^4$CFs. This is a very good description in the limit $d\gg\ell_B$. 

We see that as $d$ increases, the best overlap of the ground state is with wavefunctions where more and more fluxes bind to the electrons and holes. This makes sense because the flux binding minimizes the intralayer Coulomb repulsion. While the intralayer Coulomb repulsion is minimized via flux attachment and hence favours large $p$, the interlayer Coulomb repulsion is minimized via pairing of oppositely charged particles. Once we have attached the maximum amount of fluxes, the particles are charge neutral and the interlayer pairing does not help minimize the Coulomb energy. Hence the interlayer interaction favours small $p$. To make this argument more quantitative, we can consider the form of the potential in a planar geometry. The Fourier transform of the intralayer Coulomb interaction between $^p$CBs/$^p$CFs is $V_{\uparrow\uparrow}(q)=V_{\downarrow\downarrow}(q)=e^{*2}/(2\epsilon q)$, where the effective charge of the composite particles is $e^*=e(1-p/4)$. Due to the finite interlayer separation, the interlayer Coulomb interaction is $V_{\uparrow\downarrow}(q)=e^2/(2\epsilon q)e^{-qd}$. Note that the full electron charge $e$ enters, since the Jastrow factor is only intralayer. We now consider excitons formed out of a composite particle in the top layer and an anti composite particle in the bottom layer. Using a similar argument as in \cite{ShouCheng}, we find that interaction potential between two such excitons is
\begin{eqnarray}
    V(q)&=&V_{\uparrow\uparrow}(q)+V_{\downarrow\downarrow}(q)-2V_{\uparrow\downarrow}(q)\\
    &=&2V_{\uparrow\uparrow}(q)(q)[(1-p/4)^2-e^{-qd}]
\end{eqnarray}
A typical scale for the momentum is $q=1/\ell_B$, for which the interaction between excitons vanishes at a critical distance
\begin{equation}
    \frac{d_c(p)}{\ell_B}=-2\ln{(1-p/4)}=0, 0.58, 1.39, 2.77, \infty
\end{equation}
for $p=0,1,2,3,4$. The trial wavefunctions will be most effective when the exciton interaction energy vanishes, since this is an energetically favourable configuration. Indeed, the sequence of $d_c(p)$  roughly matches the regions where the overlap of a given trial state is maximal. Given the crudeness of the approximation, we do not expect more quantitative agreement. 

One additional aspect is that the particles with four fluxes attached feel no magnetic field and hence their wavefunctions are plane waves, the single-particle orbitals we use to construct the trial state are completely delocalized and hence we cannot use them to build excitons. (Note that for the $(\nu_\uparrow,\nu_\downarrow)=(\frac{1}{2},\frac{1}{2})$ bilayer studied in \cite{Wagner2021,Hu2024} the $^2$CFs do experience a residual net flux due the particular shift chosen to obtain the Halperin (111) state in the sphere geometry and hence this argument is not applicable there.)  The spread of the orbitals is given by the effective magnetic length $\ell_{B}^\mathrm{eff}=1/\sqrt{eB(1-p/4)}$. The interlayer Coulomb energy favours small $\ell_{B}^\mathrm{eff}$ and this effect likely further increases $d_c$.

\section{Excited state trial wavefunctions}
\label{sec:excitations}

\begin{figure*}
    \centering
    \includegraphics[width=\textwidth]{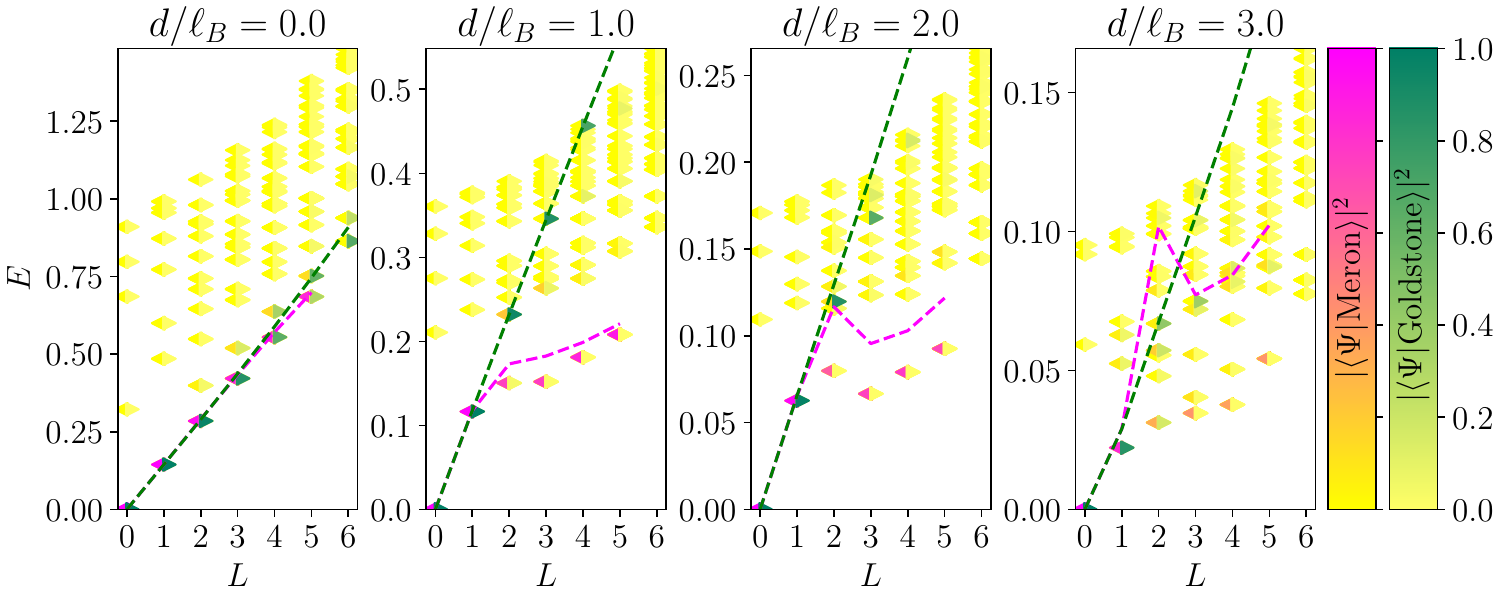}
    \caption{\textbf{Excitation spectrum and overlaps of the excited state trial wavefunctions. } We show the excitation spectrum for four values of the interlayer separation for system size $N_\uparrow=5$. We show the overlap of the two trial states, the Goldstone mode Eq.~\eqref{eq:Goldstone} and the meron mode Eq.~\eqref{eq:Merons}, with the exact diagonalization excited states. For each state in the spectrum, the left triangle shows the overlap with the meron mode (pink corresponding to high overlap) and the right triangle shows the overlap with the Goldstone mode (green corresponding to high overlap). The dashed lines show the energy expectation values of the two trial states. At $d=0$, the lowest excitation is well-described by the Goldstone mode. At intermediate distances $d\sim 1-2\ell_B$, the Goldstone mode persists as a higher energy excitation, however the lowest energy excitation is the meron excitation. }
    \label{fig:excitations}
\end{figure*}

We now consider excitations of the system at different interlayer distances. At $d=0$, we have the Halperin (111) state which spontaneously breaks the layer pseudospin $SU(2)$ symmetry and there is an associated gapless Goldstone mode.
The Goldstone mode we find corresponds to breaking the $U(1)$ symmetry from pseudo spin rotation in the $XY$ plane with $\langle S_z\rangle=N_\uparrow - N_\downarrow$ fixed, since we have no tunneling between layers. The Goldstone mode of the $U(1)$ symmetry survives even with finite but small interlayer distances. 

Following Ref.~\cite{Simon1} we can write down a trial wavefunction for the Goldstone mode by boosting the angular momentum of one layer with respect to the other. Given the exact diagonalization ground state $|\Psi_\mathrm{GS}\rangle$ in the $L=0$ sector, we can write down the trial state with angular momentum $L$ as 
\begin{equation}
    \label{eq:Goldstone}
    |\mathrm{Goldstone}\rangle\propto (\hat L_{+\uparrow})^L|\Psi_\mathrm{GS}\rangle,
\end{equation}
where the operator $\hat L_{+\uparrow}$ raises the angular momentum of the electrons in the upper layer. As shown in Fig.~\ref{fig:excitations}, at small distances the lowest energy excitations have high overlap with the trial state Eq.~\eqref{eq:Goldstone} (green triangles) and the dispersion of the Goldstone mode is linear. At $d=0$, we have tightly bound electron-hole excitons (i.e.~an exciton condensate). At $d/\ell_B=1$, we still have the tightly bound charged $^2$CFs and anti-$^2$CFs (i.e.~a \textit{composite exciton} condensate). For the ground states with the exciton or composite exciton condensate, we have the Goldstone mode of the $U(1)$ symmetry breaking. At $d/\ell_B \gtrsim 3$, $^4$CFs and anti-$^4$CFs are both neutral, we don't expect a composite exciton condensate. Instead, if we have two decoupled Fermi liquids of $^4$CFs, the lowest energy excitations of each individual layer will be particle-hole excitations of the composite Fermi liquid. Furthermore, due to the weak interlayer correlations, boosting the angular momentum of one layer with respect to the other also yields a low energy excitation. This explains why the `Goldstone mode' reappears as a low energy excitation at $d/\ell_B=3$ as was also seen in \cite{Simon1}.

At $d/\ell_B\sim 1$ we see the emergence of a new low-energy excitation that is separated by a gap from the rest of the spectrum. This excitation branch terminates at $L=N_\uparrow$. At these interlayer separations, we saw good overlap with the bosonic condensate of $^1$CBs (Fig.~\ref{fig:overlaps}). The low-energy excitations of this state are merons \cite{Moon_Review}. We can write down a trial wavefunction for this state at momentum $L$ given the exact diagonalization ground state $|\Psi_\mathrm{GS}\rangle$ as  
\begin{equation}
     \label{eq:Merons}
    |\mathrm{Meron}\rangle\propto \sum_{\{i_1\dots i_J\}}\hat L_{+\uparrow i_1}\dots \hat L_{+\uparrow i_L}|\Psi_\mathrm{GS}\rangle,   
\end{equation}
where $\hat L_{+\uparrow i}$ raises the angular momentum of the electron with index $i$. The sum runs over the partitions with $L$ elements out of the $N_1$ electrons in the top layer. This explains why the mode terminates at $L=N_\uparrow$. For $L=1$ the modes Eq.~\eqref{eq:Goldstone} and Eq.~\eqref{eq:Merons} coincide. The trial state Eq.~\eqref{eq:Merons} has very high overlap with the exact diagonalization excitation state at $1\lesssim d/\ell_B\lesssim 2$ (pink triangles in Fig.~\ref{fig:excitations}). Similar low-lying states (in particular analogous to Eq.~\eqref{eq:Merons} for $L=N_\uparrow$) have been seen in the torus geometry in \cite{Milovanovic,Park1} and they can be shown to be exact zero-energy states for a hard-core interaction \cite{MacDonald1996}. Similar excitations can also be seen for quantum Hall edges \cite{Palacios1996}.

\section{Discussion}
\label{sec:discussion}

We investigated a trial wavefunction for quantum Hall bilayers at filling $\nu=\frac{1}{4}+\frac{3}{4}$. At small distances, the quantum Hall bilayer is an exciton condensate of interlayer electron-hole pairs. At larger distances, the electrons and holes in the two layers each bind a vortex and an anti-vortex to form a $^1$CB and anti-$^1$CB pair. The system at $d/\ell_B\sim1$ can be well-described as a condensate of these $^1$CB-excitons. At yet larger distances, the electrons and holes each bind an additional vortex to form a $^2$CF and anti-$^2$CF pair.  The system at $d/\ell_B\sim 2$ can be well-described in terms of these $^2$CF-excitons. Next, an additional vortex binding yields $^3$CB-excitons, which describe the system best around  $d/\ell_B\sim 3$. A final vortex binding yields $^4$CFs which experience no net magnetic field. The CF excitons yield the best description of the system at the largest interlayer distances $d$ and capture the transition of the system to decoupled composite Fermi liquids in the two layers.

We can also understand the excitation spectrum by constructing two trial states, one for the Goldstone mode and one for a meron states. At small interlayer separations, the lowest energy excitation has good overlap with the Goldstone mode, while at intermediate interlayer separation, the lowest energy mode has good overlap with the meron state. 

The case of imbalanced quantum bilayers at total filling factor $\nu=1$ has been studied experimentally in Ref.~\cite{ImbalanceExp}. In particular, the experiments covers a range of imbalances between $\nu=\frac{1}{2}+\frac{1}{2}$ and $\nu=\frac{1}{4}+\frac{3}{4}$. There are two competing composite fermion pictures at the imbalanced configurations: $^2$CF/anti-$^2$CF pairing and $^4$CF/anti-$^4$CF pairing. One may expect to find a transition between these pictures as the imbalance is tuned. Such a transition is indeed suggested by the tunnelling conductance at various temperatures in the charge imbalanced configurations~\cite{ImbalanceExp}. We leave a more detailed comparison to experiment to future work.

Recent work has proposed a picture of CFs as Dirac fermions \cite{Son,Dirac-PH}. The CF-CF pairing of Dirac composite fermions was used to investigate the balanced case $\nu=\frac{1}{2}+\frac{1}{2}$ \cite{Sodemann}. In future work, it would be interesting to study composite fermion-composite hole pairing of the generalized Dirac composite fermion \cite{Dirac14} to describe the bilayer quantum Hall system at filling fraction $\nu=\frac{1}{4}+\frac{3}{4}$.\\

\section*{Acknowledgements}

G.W.~acknowledges funding from the University of Zurich postdoc grant FK-23-134. G.W.~would like to thank the Institute for Basic Science in Daejeon, South Korea, where part of this work was completed. D.X.N.~is supported by Grant No.~IBS-R024-D1. The exact diagonalization calculations were performed using DiagHam.

 	\bibliography{bib.bib}

\clearpage 
\newpage

\onecolumngrid
	\begin{center}
		\textbf{\large --- Supplementary Material ---\\ Successive electron-vortex binding in quantum Hall bilayers at $\nu=\frac{1}{4}+\frac{3}{4}$}\\
		\medskip
		\text{Glenn Wagner and Dung X. Nguyen}
	\end{center}
	
		\setcounter{equation}{0}
	\setcounter{figure}{0}
	\setcounter{table}{0}
	\setcounter{page}{1}
	\makeatletter
	\renewcommand{\theequation}{S\arabic{equation}}
	\renewcommand{\thefigure}{S\arabic{figure}}
	\renewcommand{\bibnumfmt}[1]{[S#1]}
\begin{appendix}

\section{Details of lowest Landau level projection}

The trial wavefunction we wrote in the main text is
\begin{align}
    \psi_p&=\mathcal{P}_\textrm{LLL}\bigg[\Pi_{i<j}(\Omega_i-\Omega_j)^p(\varpi^*_i-\varpi_j^*)^p\ f\bigg( \sum_{n=0}^{N_\mathrm{max}}g_n\sum_{m=-n}^{n}  Y_{qnm}(\Omega_i) Y_{qnm}(\varpi_j)^*\bigg)\bigg]\\
    &=\mathcal{P}_\textrm{LLL}\ f\bigg( \sum_{n=0}^{N_\mathrm{max}}g_n\sum_{m=-n}^{n} [\Pi_{k\neq i}(\Omega_k-\Omega_i)^p  Y_{qnm}(\Omega_i)][ \Pi_{l\neq j}(\varpi_l-\varpi_j)^p Y_{qnm}(\varpi_j)]^*\bigg)
\end{align}
Since $Y_{qnm}(\Omega)$ does not generally lie in the lowest Landau level, we apply the Jain-Kamilla \cite{KamillaJain} projection procedure to the monopole harmonics. The Jain-Kamilla projection procedure is usually employed for composite fermion orbitals, but it has been carried out in \cite{Balram2020} for composite boson orbitals. The Jain-Kamilla projection is performed by letting $\mathcal{P}_\textrm{LLL}$ act on different terms inside the wavefunction separately. This is an approximation, which however can be shown to lead to trial states that have excellent overlap with the exact diagonalization ground states. To wit, we use 
\begin{equation}
    \mathcal{P}_\textrm{LLL}\ \Pi_{k\neq i}(\Omega_k-\Omega_i)^\beta Y_{qnm}(\Omega_i)=\Pi_{k\neq i}(\Omega_k-\Omega_i)^\beta \tilde Y_{qnm}(\Omega_i)
\end{equation}
where $\tilde Y_{qnm}(\Omega)$ are the Jain-Kamilla orbitals, for which explicit expressions can be found in \cite{KamillaJain}. $\beta$ should be even and we pick $\beta=2,2,2,4$ for $p=1,2,3,4$ respectively. With this, we end up with the final expression for the lowest Landau level projected trial state:
\begin{equation}
    \psi_p=(\Omega_i-\Omega_j)^p(\varpi^*_i-\varpi_j^*)^p\ f\bigg(\sum_{n=0}^{N_\mathrm{max}}g_n\sum_{m=-n}^{n} \tilde Y_{qnm}(\Omega_i)\tilde Y_{qnm}(\varpi_j)^*\bigg).
\end{equation}

\section{Additional numerical results}

\begin{figure}[h]
    \centering
    \includegraphics[width=0.9\textwidth]{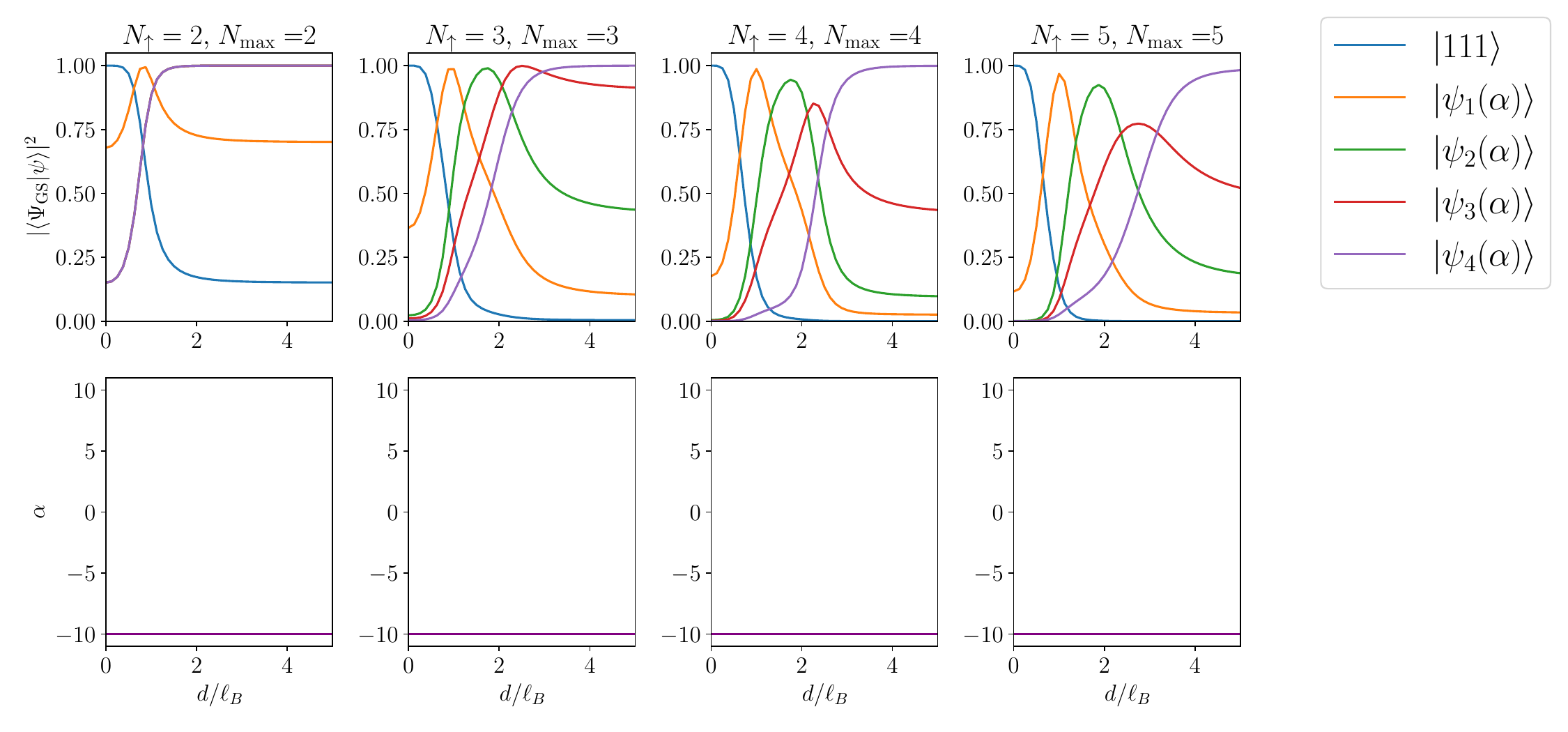}
    \caption{Same figure as Fig.~\ref{fig:overlaps} without optimizing the variational parameters. We fix $\alpha=-10$ for all variational wavefunctions.}
    \label{fig:enter-label}
\end{figure}

\begin{figure}[h]
    \centering
    \includegraphics[width=\textwidth]{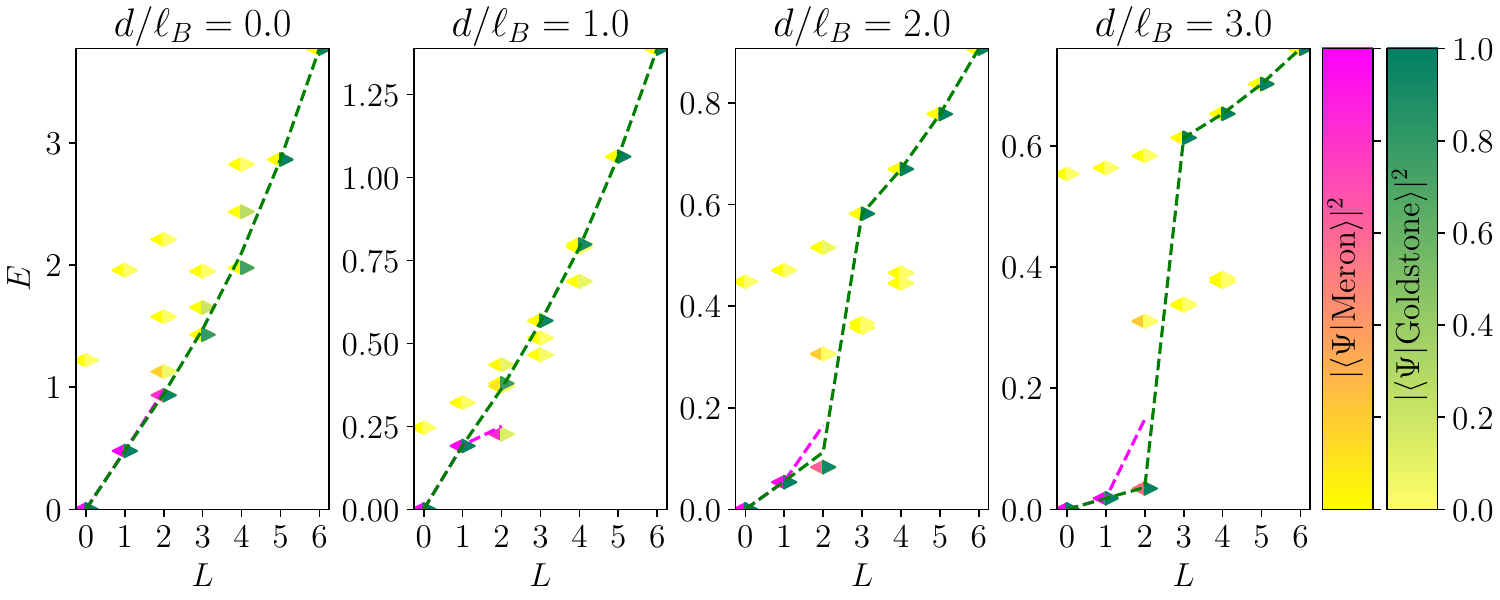}
    \caption{Same figure as Fig.~\ref{fig:excitations} but for $N_\uparrow=2$. The meron mode terminates at $L=2$.}
    \label{fig:enter-label}
\end{figure}

\begin{figure}[h]
    \centering
    \includegraphics[width=\textwidth]{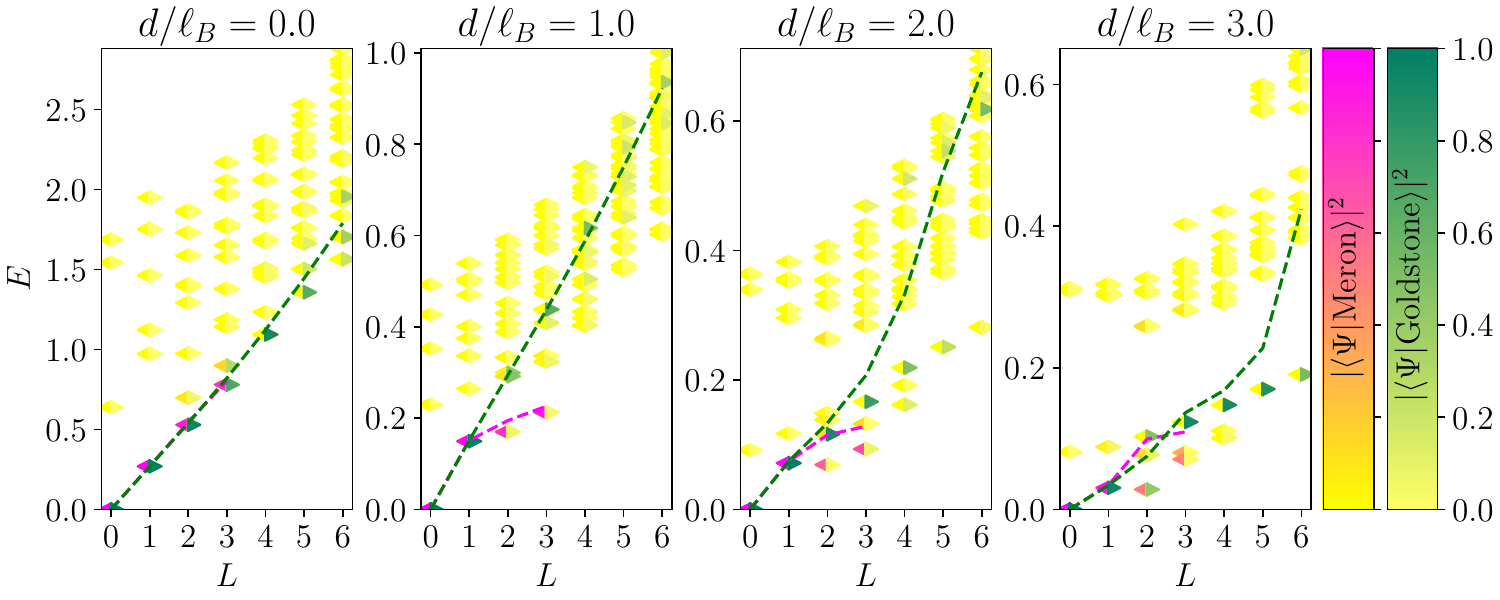}
    \caption{Same figure as Fig.~\ref{fig:excitations} but for $N_\uparrow=3$. The meron mode terminates at $L=3$.}
    \label{fig:enter-label}
\end{figure}

\begin{figure}[h]
    \centering
    \includegraphics[width=\textwidth]{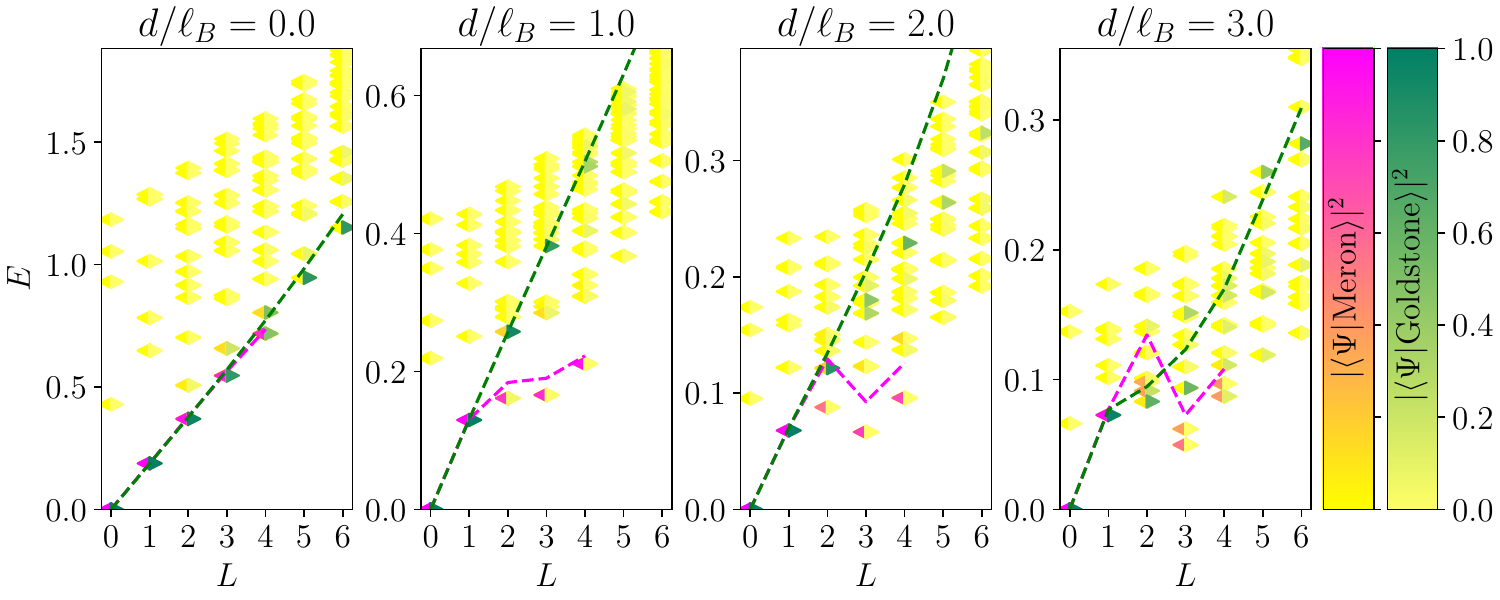}
    \caption{Same figure as Fig.~\ref{fig:excitations} but for $N_\uparrow=4$. The meron mode terminates at $L=4$.}
    \label{fig:enter-label}
\end{figure}

\end{appendix}
\end{document}